\documentclass[reprint,amssymb,amsmath,floatfix,aps,prd,groupedaddress,nofootinbib]{revtex4-2}
\usepackage{bbm} 
\usepackage[caption=false]{subfig}
\usepackage{graphicx}
\usepackage{hyperref}
\usepackage{bm}
\usepackage{color}
\usepackage{slashed}
\allowdisplaybreaks

\newcommand{\tr}[1]{\left\langle #1 \right\rangle}
\newcommand{\com}[2]{\left[ #1,#2 \right]}

\newcommand{\sdot}[2]{\left( #1\cdot #2 \right)}

\begin{document}
\title{Pion axioproduction:
The Delta resonance contribution}

\author{Thomas Vonk}
 \email{vonk@hiskp.uni-bonn.de}
 \affiliation{Helmholtz-Institut f\"{u}r Strahlen- und Kernphysik and Bethe Center for Theoretical Physics,
   Universit\"{a}t Bonn, D-53115 Bonn, Germany}
 
 \author{Feng-Kun Guo}
 \email{fkguo@itp.ac.cn}
 \affiliation{CAS Key Laboratory of Theoretical Physics, Institute of Theoretical Physics, Chinese Academy
   of Sciences, Beijing 100190, China}
 \affiliation{School of Physical Sciences, University of Chinese Academy of Sciences,
    Beijing 100049, China}

\author{Ulf-G. Mei{\ss}ner}
 \email{meissner@hiskp.uni-bonn.de}
 \affiliation{Helmholtz-Institut f\"{u}r Strahlen- und Kernphysik and Bethe Center for Theoretical Physics,
   Universit\"{a}t Bonn, D-53115 Bonn, Germany}
 \affiliation{Institute for Advanced Simulation, Institut f\"ur Kernphysik and J\"ulich Center for Hadron Physics,  Forschungszentrum J\"ulich, D-52425 J\"ulich, Germany}
 \affiliation{Tbilisi State University, 0186 Tbilisi, Georgia}


\begin{abstract}
The process of pion axioproduction, $aN\to\pi N$, with an intermediate $\Delta$ resonance is analyzed
using baryon chiral parturbation theory. The $\Delta$ resonance is included in two ways:
First, deriving the $a\Delta N$-vertices, the axion is brought into contact with the resonance, and,
second, taking the results of $\pi N$ elastic scattering including the $\Delta$, it is implicitly included in
the form of a pion rescattering diagram. As a result, the partial wave cross section of axion-nucleon
scattering shows an enhancement in the energy region around the $\Delta$ resonance. Because of the isospin breaking, the enhancement is not as pronounced as previously anticipated.
However, since the isospin breaking here is much milder than that for usual hadronic processes, novel axion search experiments might still exploit this effect.

\end{abstract}

\maketitle

\section{Introduction}

A model that might resolve two of the known problems of two different (but related) physical fields, in the present 
case the strong-CP problem of quantum chromodynamics (QCD) and the dark matter issue of astrophysics and cosmology
\cite{Preskill:1982cy,Abbott:1982af,Dine:1982ah,Ipser:1983mw,Turner:1986tb,Duffy:2009ig,Marsh:2015xka}, is clearly
worth investigating. Such a model is the Peccei--Quinn model~\cite{Peccei:1977hh,Peccei:1977ur} and the theory
of the axion~\cite{Weinberg:1977ma,Wilczek:1977pj},  especially the ``invisble'' axion models such as the
Kim--Shifman--Vainstein--Zakharov (KSVZ) axion model~\cite{Kim:1979if,Shifman:1979if} and the
Dine--Fischler--Srednicki--Zhitnitsky (DFSZ) axion model~\cite{Dine:1981rt,Zhitnitsky:1980tq}. However,
the time theorists and experimentalists effortfully spent on the search for signals that might verify this
model now comprises more than four decades and still there is no axion in sight. Because of that, it is important
to study all kinds of related processes hoping to figure out some underlying phenomenon that might enhance
the chance of its detection, if only for a few percent.

Recently, Carenza et al.~\cite{Carenza:2020cis} have proposed that the pion axioproduction\footnote{We use the term axioproduction in analogy to terms like electroproduction or photoproduction. Pion axioproduction hence means pion production induced by axions.} $aN\to \pi N$ is such a process, 
because at certain axion energies, around 200--300~MeV, an enhanced axion-nucleon cross section due to the $\Delta$ resonance
can be expected. This in turn would possibly make axion detections accessible for underground water Cherenkov
detectors. Such axions might be produced in protosupernova cores in the presence of pions, where besides
the axion production via axion-nucleon Bremsstrahlung $NN\to aNN$, the pion-induced process $ \pi N \to aN$
might play a more important role than previously thought~\cite{Carenza:2020cis,Fischer:2021jfm}, leading to a possible enhancement of the number spectrum of axions with energies around 200--300~MeV. 

In this study, we take a closer look at exactly this process, namely $aN\to \pi N$ with the $\Delta$ resonance, showing
that there is indeed a region of enhancement. This enhancement is, however, by at least an order of magnitude
less pronounced than that anticipated by Carenza et al.~\cite{Carenza:2020cis}, which we will discuss in more detail
below in Sec.~\ref{sec:result}.

Having said that, it is important to remind the reader that the traditional window for the QCD axion as a
dark matter candidate dictates~\cite{Preskill:1982cy,Abbott:1982af,Kim:1986ax,Kim:2008hd}
\begin{equation}\label{eq:window}
10^{9}\,\text{GeV} \lesssim f_a \lesssim 10^{12}\,\text{GeV}\,,
\end{equation}
where $f_a$ is the axion decay constant which eventually controls and suppresses the axion
mass~\cite{Lu:2020rhp,DiLuzio:2020wdo}
\begin{equation}
m_a \approx 5.7 \left(\frac{10^{12}\,\text{GeV}}{f_a} \right) \times 10^{-6}  \,\text{eV}
\end{equation}
and the axion-nucleon coupling $G_{aN} \propto 1/f_a$. This means that despite the possible enhancement due to the presence of baryon resonances, the reaction cross section still remains tiny.

A very suitable framework for studying the process at hand is chiral perturbation theory (CHPT), which
has been successfully extended to the meson-nucleon and $\Delta$-meson-nucleon sectors, and which in the
past also has been applied to the study of the axion-nucleon interaction~\cite{GrillidiCortona:2015jxo,Vonk:2020zfh},
after the leading order axion-nucleon interaction had been studied for years in the context of current
algebra, which is equivalent to a leading order calculation in CHPT~\cite{Donnelly:1978ty,Kaplan:1985dv,Srednicki:1985xd,Georgi:1986df,Chang:1993gm}. In this paper
we use these results including the thereby accrued knowledge of the underlying structure of the
axion-nucleon coupling. Moreover, we show how to include the $\Delta$ baryon into the model. 

In Sec.~\ref{sec:basic} we first give a short discussion of the kinematics and the general isospin
structure of the $aN\to \pi N$ scattering amplitude, as well as a brief presentation of baryon CHPT with axions and the  $\Delta$ resonance. After that, we  work out the amplitudes of the individual
Feynman diagrams contributing to the pion axioproduction in Sec.~\ref{sec:amplitudes}. Putting pieces
together, we finally discuss the results in Sec.~\ref{sec:result}.

\section{Theoretical foundation}\label{sec:basic}
\subsection{Kinematics}
The process under consideration is
\begin{equation}
a(q) + N(p)  \to \pi^b(q^\prime) + N(p^\prime) ,
\end{equation}
where $a$ denotes the axion, $N$ a nucleon, either the proton or the neutron, and $\pi^b$ a pion with the isospin index $b$.
As usual, we define the Lorentz invariant Mandelstam variables
\begin{equation}\label{eq:mandelstam}
s = (p+q)^2,\qquad t = (p-p^\prime)^2,\qquad u = (p-q^\prime)^2
\end{equation}
for the four-momenta $q,p$ of the incoming particles and $q^\prime,p^\prime$ of the outgoing particles. The
invariants of Eq.~\eqref{eq:mandelstam} fulfill the on-shell relation
\begin{equation}
s+t+u = 2m_N^2 + m_a^2 + M_\pi^2 , \label{eq:MandelstamIdentity}
\end{equation}
which can be used to eliminate one of the three variables, which we choose to be $u$. Throughout this paper
we use the center-of-momentum (c.m.) system, where for the three-momenta $\mathbf{p}+\mathbf{q}=
\mathbf{p^\prime}+\mathbf{q^\prime}=0$. Using the well-known K\"all\'{e}n function
\begin{equation}
\lambda(a,b,c) = a^2+b^2+c^2-2ab-2ac-2bc ,
\end{equation}
the c.m. energies of the incoming and outgoing nucleons can be written as
\begin{equation}
E_\mathbf{p} = \frac{s+m_N^2-m_a^2}{2\sqrt{s}},\qquad E_\mathbf{p^\prime} = \frac{s+m_N^2-M_\pi^2}{2\sqrt{s}},
\end{equation}
and one has
\begin{align}
\begin{split}
|\mathbf{p}|=|\mathbf{q}| & =\frac{\sqrt{\lambda(s,m_N^2,m_a^2)}}{2\sqrt{s}}, \\
 |\mathbf{p^\prime}|=|\mathbf{q^\prime}| & =\frac{\sqrt{\lambda(s,m_N^2,M_\pi^2)}}{2\sqrt{s}} .
\end{split}
\end{align}
Moreover, setting $z=\cos\theta$, where $\theta$ is the c.m. scattering angle, we have
\begin{equation}
\sdot{\mathbf{p}}{\mathbf{p^\prime}}=|\mathbf{p}||\mathbf{p^\prime}| z ,
\end{equation}
so we can reexpress the second Mandelstam $t$ variable as
\begin{equation}
t = 2 \left(m_N^2 - E_\mathbf{p} E_\mathbf{p^\prime} + |\mathbf{p}||\mathbf{p^\prime}| z \right) .
\end{equation}
Before discussing how these kinematic quantities enter the scattering amplitudes, we briefly take a look at
the isospin structure of the process.

\subsection{Isospin structure}\label{ssec:isospin}

For the $\pi N$ elastic scattering, it is common to decompose the scattering amplitude $T^{ab}_{\pi N\to\pi N}$, where $a$ is
the isospin index of the incoming pion and $b$ for the outgoing one, according to the isospin structure.
In the isospin limit, the decomposition reads
\begin{equation}\label{eq:piNiso}
T^{ab}_{\pi N\to\pi N} = T^+ \delta_{ab} + T^- \frac{1}{2}\com{\tau_a}{\tau_b},
\end{equation}
where $\tau_a$ and $\tau_b$ are the Pauli matrices and $\com{\ }{\ }$ denotes the commutator. However, for the
present process $aN\to \pi^b N$ we are particularly interested in transitions including the $\Delta$ resonance, as suggested in Ref.~\cite{Carenza:2020cis},
which is an isospin-$\tfrac{3}{2}$ particle. As the axion is an isoscalar, no isospin symmetric $aN$
interaction can lead to the appearance of the $\Delta$ resonance, so we are especially interested in isospin
breaking interactions. Indeed, the isovector axial-vector current $a_\mu$ (see below) introduces such
isospin breaking pieces into the axion-baryon interaction, which can be seen, for instance, below in
Eqs.~\eqref{eq:cuds} and \eqref{eq:aNcoupling}.

As it turns out, it is possible to decompose the scattering amplitude $T^{b}_{aN\to\pi N}$ into
\begin{equation}\label{eq:isoprojection}
T^{b}_{aN\to\pi N} =  T^+ \delta_{3b} +  T^{3+} \tau_3 + T^{-} \frac{1}{2}\com{\tau_b}{\tau_3}  ,
\end{equation}
which is comparable to the case of $\pi N$ scattering with isospin violation, see, e.g.,
Refs.~\cite{Fettes:2000vm,Hoferichter:2009gn}. Any of the four possible amplitudes can be expressed by means
of the three objects $ T^+, T^{3+}$, and $T^{-}$:
\begin{align}
\begin{split}
T_{ap\to\pi^0 p} & = T^+ + T^{3+}\,, \\ 
T_{an\to\pi^0 n} & = T^+ - T^{3+}\,, \\
T_{ap\to\pi^+ n} & = \sqrt{2}\left(T^{3+} + T^{-}\right),\\
T_{an\to\pi^- p} & = \sqrt{2}\left(T^{3+} - T^{-}\right).
\end{split}
\end{align}
The part of the amplitude that leads to isospin violation and thus to possible enhancement due to the $\Delta$
resonances, which we denote by $T^{3/2}$, is found by taking the difference
\begin{equation}\label{eq:Tdifference}
T^{3/2} = T^+ - T^{-}\,,
\end{equation}
or alternatively,
\begin{align}\label{eq:Tdifference2}
\begin{split}
T^{3/2}	& =  T_{ap\to\pi^0 p} - \frac{1}{\sqrt{2}} T_{ap\to\pi^+ n} \\
  	& =  T_{an\to\pi^0 n} + \frac{1}{\sqrt{2}} T_{an\to\pi^- p}\,,
\end{split}
\end{align}
where the latter expressions have the advantage that one can also easily account
for differences in the charged and neutral pion masses, which improves the accuracy of the calculation.

\subsection{Partial wave decomposition}

It is known from  $\pi N$ scattering that the $\Delta$ resonance chiefly affects the $P_{33}$ partial wave
(where we, as usual, make use of the spectroscopic notation $l_{2I,2j}$, $l = S, P, D, \dots$ being the
orbital angular momentum, $I$ the isospin, and $j=l+s$ the total angular momentum). Therefore, it
is expedient to focus on the $P_{33}$ partial wave also in the present study of the $aN\to\pi N$ reaction.

To this end, we decompose any of the amplitudes given above as
\begin{equation}\label{eq:tdecomp}
T = \bar{u}(p^\prime) \left\{A(s,t) + B(s,t)\frac{1}{2}\left(\slashed{q} + \slashed{q}^\prime\right)\, \right\} u(p) ,
\end{equation}
where we make use of the well-known notation, $\slashed{q} = \gamma^\mu q_\mu$. One then can project out
any partial wave of definite total angular momentum $j=l\pm 1/2$, abbreviated as $l\pm$, by
\begin{widetext}
\begin{align}
\begin{split}
T_{l\pm} (s) = & \frac{\sqrt{E_\mathbf{p}+m_N}\sqrt{E_\mathbf{p^\prime}+m_N}}{2}\left\{A_l(s) + \left(\sqrt{s}-m_N\right) B_l(s) \right\} \\
 & + \frac{\sqrt{E_\mathbf{p}-m_N}\sqrt{E_\mathbf{p^\prime}-m_N}}{2}\left\{-A_{l\pm 1}(s) + \left(\sqrt{s}+m_N\right) B_{l\pm 1}(s) \right\},
\end{split}\label{eq:lprojection}
\end{align}
\end{widetext}
where
\begin{align}
\begin{split}
A_l (s) & = \int_{-1}^{+1} A(s,t(s,z)) P_l(z)\ \text{d}z,\\ B_l (s) & = \int_{-1}^{+1} B(s,t(s,z)) P_l(z)\ \text{d}z
\end{split}
\end{align}
using the well-known Legendre polynomials $P_l(z)$.

\subsection{Partial wave cross section}

For experiments, the most useful quantity is the cross section
\begin{equation}
\mathrm{d}\sigma = \frac{1}{\mathcal{F}}|\mathcal{M}|^2 \mathrm{d}\Pi_2\,,
\end{equation}
with the flux factor
\begin{equation}
\mathcal{F} = 4 \sqrt{\sdot{p}{q}-m_N m_a} = 4 |\mathbf{p}| \sqrt{s}
\end{equation}
and the two-body phase space
\begin{align}
\int \mathrm{d}\Pi_2 &= \int \frac{\mathrm{d}^3p^\prime}{(2\pi)^3} \frac{\mathrm{d}^3q^\prime}{(2\pi)^3}
\frac{1}{2E_\mathbf{p^\prime}2E_\mathbf{q^\prime}} (2\pi)^4 \delta^4(p+q-p^\prime-q^\prime) \nonumber\\
&=
\int \mathrm{d}\Omega \frac{1}{16\pi^2} \frac{|\mathbf{p}^\prime|}{\sqrt{s}} ,
\end{align}
where in both cases the right-most expressions are valid in the c.m. frame. The total cross section
is hence given by
\begin{equation}
\sigma = \frac{1}{64\pi^2s}\frac{|\mathbf{p^\prime}|}{|\mathbf{p}|}\int\mathrm{d}\Omega\,|\mathcal{M}|^2 ,
\end{equation}
which can be expanded in terms of partial wave cross sections as
\begin{equation}
\sigma = \sum_{l} \sigma_{l\pm}\,.
\end{equation}
The inverse of Eqs.~\eqref{eq:lprojection} is given by
\begin{widetext}\begin{equation}
T = 2m_N \chi_f^\dagger  \sum_l \left\{ \left[(l+1)\, T_{l+} +l\,T_{l-} \right] P_l (z) -
i\,\bm{\sigma}\cdot (\mathbf{\hat{q}^{\prime}}\times\mathbf{\hat{q}}) (T_{l+} - T_{l-})
\frac{\mathrm{d}P_l}{\mathrm{d}z} \right\}\chi_i\,,
\end{equation}
\end{widetext}
where $\chi_i$ and $\chi_f$ are the Pauli spinors of the incoming and outgoing nucleons, respectively,
and $\mathbf{\hat{q}^{(\prime)}} = \mathbf{q^{(\prime)}}/|\mathbf{q^{(\prime)}}|$. For the $j=\tfrac{3}{2}$ case, one finds
\begin{equation}\label{eq:pwcrosssection}
\sigma_{1+} = \frac{1}{8\pi s} \frac{|\mathbf{p^\prime}|}{|\mathbf{p}|} |T_{1+}|^2\,.
\end{equation}

The bottom line of the previous elaborations then is that we will derive the amplitudes $A^{\pm,3+}(s,t)$
and $B^{\pm,3+}(s,t)$ for any Feynman diagram of interest and use Eqs.~\eqref{eq:isoprojection}
and~\eqref{eq:lprojection} in order to determine the $P_{33}$ partial wave amplitude $T^{33}_{aN\to\pi N}$
for the pertinent processes. This amplitude in turn is used to ascertain the corresponding cross section
via Eq.~\eqref{eq:pwcrosssection}. The theoretical framework of determining $A^{\pm,3+}(s,t)$ and $B^{\pm,3+}(s,t)$
is CHPT.

\subsection{Baryon chiral perturbation theory with axions}

The way of incorporating the axion into CHPT is discussed in detail in Ref.~\cite{Vonk:2020zfh}. Here, we will
only outline the major steps. First, recall that in the standard QCD axion models, the KSVZ and DFSZ ones,
the axion-quark couplings $X_q$ appearing in the QCD Lagrangian after the spontaneous breakdown of
Peccei--Quinn symmetry are flavor-diagonal and given by 
\begin{align}
\label{eq:couplignconstantsmodeldepending}
\begin{split}
X_q^\mathrm{KSVZ} & = 0 \, , \\
X_{u,c,t}^\mathrm{DFSZ} & = \dfrac{1}{3} \dfrac{x^{-1}}{x+x^{-1}} = \dfrac{1}{3}\sin^2\beta\, , \\
X_{d,s,b}^\mathrm{DFSZ} &=\dfrac{1}{3} \dfrac{x}{x+x^{-1}}=\dfrac{1}{3}\cos^2\beta = \dfrac{1}{3}
- X_{u,c,t}^\mathrm{DFSZ} \, , 
\end{split}
\end{align}
where $x = \cot \beta$ is the ratio of the vacuum expectation values of the two Higgs doublets in the
DFSZ model. After a chiral rotation removing the axion-gluon coupling terms in the Lagrangian, the whole axion-quark
interaction can be decomposed into isovector and isoscalar parts with  the couplings
\begin{align}
\label{eq:cuds}
\begin{split}
c_{u-d} &= \frac{1}{2}\left(X_u-X_d-\frac{1-z}{1+z+w}\right)  ,  \\
c_{u+d} &= \frac{1}{2}\left(X_u+X_d-\frac{1+z}{1+z+w}\right)  , \\
c_s & = X_s-\frac{w}{1+z+w}\, , \\
c_{c,b,t} & =X_{c,b,t}\,,
\end{split}
\end{align}
where $z=m_u/m_d$ and $w=m_u/m_s$ are the quark mass ratios of the three light quarks. In what follows,
the $c_i$, $i=\{1,\dots,5\}$, refer to the isoscalar couplings $\{u+d,s,c,b,t\}$ and in any  equation a
summation over repeated $i$ is implied. It is these couplings that enter the Lagrangian of CHPT in the
form of external currents\footnote{Note the typo in Ref.~\cite{Vonk:2020zfh}, Eq.~(3.16). $\tilde{u}_{\mu,i}$, which
corresponds to $u_{\mu,i}$ in the present paper, is of course not $\propto \tau_3$, but $\propto\mathbbm{1}$.}
\begin{equation}\label{eq:externalamu}
a_\mu = c_{u-d} \frac{\partial_\mu a}{2f_a} \tau_3\, , \qquad a_{\mu,i}^{(s)} = c_i \frac{\partial_\mu a}{2f_a} \mathbbm{1}\,.
\end{equation}
The transition to CHPT is phenomenologically related to the confinement of quarks and gluons into mesons and
baryons at low energies and the observation that the QCD Lagrangian is approximately invariant under the chiral symmetry SU$(N_f)_L\times$SU$(N_f)_R$, with $N_f$ the number of light quark flavors, which is spontaneously broken into the vector subgroup.
Hence, in SU(2) baryon CHPT nucleons and pions are the relevant degrees of freedom rather
than the more fundamental quarks and gluons. The application of power counting rules then leads to a
systematic perturbative description of any low energy strong interaction process, as long as the applied
Lagrangian respects all pertinent symmetries, as first worked out by Weinberg~\cite{Weinberg:1978kz}.

For the meson-nucleon sector that we are interested in, we follow the description of baryon CHPT given in Ref.~\cite{Bernard:1995dp}. The pions enter the theory in the form of a unitary $2\times 2$ matrix
\begin{equation}
u=\sqrt{U}=\exp\left(i\frac{\pi^a\tau_a}{2F_\pi}\right),
\end{equation}
where $F_\pi$ is the pion decay constant. Strictly speaking, this should be the pion decay constant in
the chiral limit, $F$, but to the order we are working on we can use the physical value.
With this unitary matrix and the external currents $a_\mu$ and $a_\mu^{(s)}$, see Eq.~\eqref{eq:externalamu},
one forms the following basic building blocks
\begin{align}
\begin{split}
u_\mu & = i\left[ u^\dagger \partial_\mu u - u \partial_\mu u^\dagger - i u^\dagger a_\mu u - i u a_\mu u^\dagger \right], \\
u_{\mu,i} & = i\left[ - i u^\dagger a_{\mu,i}^{(s)} u - i u a_{\mu,i}^{(s)} u^\dagger \right] = 2 a_{\mu,i}^{(s)}\,, \\
D_\mu & = \partial_\mu + \Gamma_\mu \\ &= \partial_\mu + \frac{1}{2} \left[ u^\dagger \partial_\mu u + u \partial_\mu u^\dagger
-i u^\dagger a_\mu u  + i u a_\mu u^\dagger \right]. 
\end{split}\label{eq:blocks}
\end{align}
Note that we only introduce and show the axial-vector currents $a_\mu$ (isovector) and $a_\mu^{(s)}$ (isoscalar)
(and not the corresponding vector currents) as these are the only external currents that are of interest
in what follows. The last object in Eq.~\eqref{eq:blocks}, $D_\mu$, is the so-called chiral covariant derivative.
At leading order, the axion only enters the model via these building blocks. At higher order, it also enters
in the form of non-derivative interactions by means of terms involving the complex phase of the quark mass matrix.

In what follows, we only need the leading order pion-nucleon Lagrangian, which is given by
\begin{equation}\label{eq:piNLagrangian}
\mathcal{L}_{\pi N} = \bar{N} \biggl\{ i \slashed{D} - m_N + \frac{g_A}{2} \slashed{u}\gamma_5 +
\frac{g_0^i}{2} \slashed{u}_i\gamma_5 \biggr\} N\, ,
\end{equation}
where $N = (p,n)^\text{T}$ is an isodoublet containing the proton and the neutron spinors, $m_N$ is the
nucleon mass in the chiral limit, and $g_A$ and the $g_0^i$'s are the axial-vector and corresponding
isoscalar coupling constants, all also in  the chiral limit. Again, to the order we are working, we can
identify these parameters with their physical values.

In Ref.~\cite{Vonk:2020zfh}, we already worked out and used the relevant vertices that can be derived
from this Lagrangian and which are also needed for the present study. Denoting the momentum of an
incoming axion with $q_\mu$ and setting $b$ as the pion isospin index, one finds for the relevant vertices:
\begin{align}
\begin{split}
aNN :& \qquad \frac{g_{aN}}{2f_a} \slashed{q} \gamma_5\,,\\
a\pi_b NN :& \qquad i\frac{c_{u-d}}{4f_a F_\pi} \slashed{q} \com{\tau_3}{\tau_b}\,.
\end{split}\label{eq:vertices}
\end{align}
The latter contact interaction is often ignored in studies of the $aN$ reaction, which are mainly based on
the former vertex, but has recently been included in the study of axion production in
supernovae~\cite{Choi:2021ign}.

The axion-nucleon coupling appearing in Eq.~\eqref{eq:vertices} is a $2\times 2$ matrix in isospin space defined as
\begin{equation}\label{eq:aNcoupling}
g_{aN} = c_{u-d} g_A \tau_3 + c_i g_0^i \mathbbm{1}\,,
\end{equation}
from which one can directly read off the couplings of the axion to the proton, $g_{ap}$, and the neutron, $g_{an}$,
respectively.

\subsection{The \texorpdfstring{$\Delta$}{Delta} resonance in chiral perturbation theory}

The free-field Lagrangian of the four $\Delta$ baryons is given by
\cite{Jenkins:1991es,Tang:1996sq,Hemmert:1997ye,Pascalutsa:2002pi,Hacker:2005fh,Krebs:2008zb}
\begin{equation}
\mathcal{L}_{\Delta} = \bar{\Delta}_\mu \Lambda^{\mu\nu}(A)\Delta_{\nu},
\end{equation}
where
\begin{equation}
\Delta_\mu = \begin{pmatrix}
\Delta^{++}_\mu \\ \Delta^{+}_\mu \\ \Delta^{0}_\mu \\ \Delta^{-}_\mu
\end{pmatrix}
\end{equation}
is the spin-$\tfrac{3}{2}$ and isospin-$\tfrac{3}{2}$ vector-spinor field, and
\begin{align}
\begin{split}
\Lambda^{\mu\nu}(A=-1)= & -\left(i\slashed{\partial}-m_\Delta \right) \\ & +i\left(\gamma^\mu\partial^\nu + \gamma^\nu\partial^\mu-\gamma^\mu\slashed{\partial}\gamma^\nu\right) \\ & - m_\Delta\gamma^\mu\gamma^\nu \,.
\end{split}
\end{align}
Here, $m_\Delta$
denotes the mass of the $\Delta$ and $A$ is a non-physical parameter that for convenience has been
set to $-1$. The propagator for the $\Delta$ with four-momentum $p^\mu$ is then given by
\begin{align}
\begin{split}
-i\frac{\slashed{p}+m_\Delta}{p^2-m_\Delta^2}\Bigl[g^{\mu\nu}-\frac{1}{3}\gamma^\mu\gamma^\nu
 & +\frac{1}{3m_\Delta}\left(p^\mu\gamma^\nu-\gamma^\mu p^\nu\right) \\ & -\frac{2}{3m_\Delta^2}p^\mu p^\nu \Bigr] .
\end{split}
\end{align}
The $\pi N\Delta$ and the $aN\Delta$ interactions are derived from the general leading order interaction
Lagrangian given by \cite{Tang:1996sq,Krebs:2009bf}
\begin{equation}\label{eq:deltainteractionlagr}
\mathcal{L}_\text{int.} = \frac{g}{2} \bar{\Delta}_{\mu,i} \left(g^{\mu\nu}+z_0\gamma^\mu\gamma^\nu \right)
\tr{\tau_i u_\nu} N + \text{h.c.}\,,
\end{equation}
where $\text{h.c.}$ stands for the Hermitian conjugate and $\tr{\ }$ denotes the trace in flavor space.
Furthermore, we make use of the isospurion representation $\Delta_{\mu,i} = \mathcal{T}_i \Delta_\mu$ with
the $2\times 4$ isospin-$\tfrac{1}{2}$-to-isospin-$\tfrac{3}{2}$ transition matrices
\cite{Tang:1996sq,Pascalutsa:2006up}
\begin{align}
\begin{split}
\mathcal{T}_1 & =\frac{1}{\sqrt{6}} \begin{pmatrix} -\sqrt{3} & 0 & 1 & 0 \\ 0 & -1 & 0 & \sqrt{3} \end{pmatrix}, \\
\mathcal{T}_2 & =\frac{-i}{\sqrt{6}} \begin{pmatrix} \sqrt{3} & 0 & 1 & 0 \\ 0 & 1 & 0 & \sqrt{3} \end{pmatrix}, \\
\mathcal{T}_3 & =\sqrt{\frac{2}{3}} \begin{pmatrix} 0 & 1 & 0 & 0 \\ 0 & 0 & 1 & 0 \end{pmatrix} ,
\end{split}
\end{align}
such that
\begin{align}
\begin{split}
\Delta_{\mu,1} & = \frac{1}{\sqrt{2}} \begin{pmatrix}\frac{1}{\sqrt{3}}\Delta^0_\mu - \Delta^{++}_\mu \\ \Delta^-_\mu -\frac{1}{\sqrt{3}} \Delta^{+}_\mu  \end{pmatrix}, \\
\Delta_{\mu,2} & = -\frac{i}{\sqrt{2}} \begin{pmatrix}\frac{1}{\sqrt{3}}\Delta^0_\mu + \Delta^{++}_\mu \\ \Delta^-_\mu +\frac{1}{\sqrt{3}} \Delta^{+}_\mu  \end{pmatrix}, \\
\Delta_{\mu,3} & = \sqrt{\frac{2}{3}} \begin{pmatrix}\Delta^+_\mu \\ \Delta^0_\mu \end{pmatrix}.
\end{split}
\end{align}
The interaction Lagrangian Eq.~\eqref{eq:deltainteractionlagr} contains two coupling constants $g$ and $z_0$, 
the latter being an off-shell parameter. $N$ again denotes the nucleon doublet and $u_\mu$ has been given
already above in Eq.~\eqref{eq:blocks}. Note that this interaction Lagrangian only allows for isovector
interactions with external axial currents $a_\mu$, whereas isoscalar interactions with $a_{\mu,i}^{(s)}$
vanish as a consequence of the trace operation. This reflects what has been said already above: any
$aN\Delta$ interaction must come with isospin violation, which is only present in 
$a_\mu$, not in $a_{\mu,i}^{(s)}$.

\section{Relevant diagrams and their contributions}\label{sec:amplitudes}
\subsection{Contact contribution and intermediate nucleon}\label{ssec:treeaN}

\begin{figure*}[t]
\subfloat[]{
\includegraphics[width=0.28\textwidth]{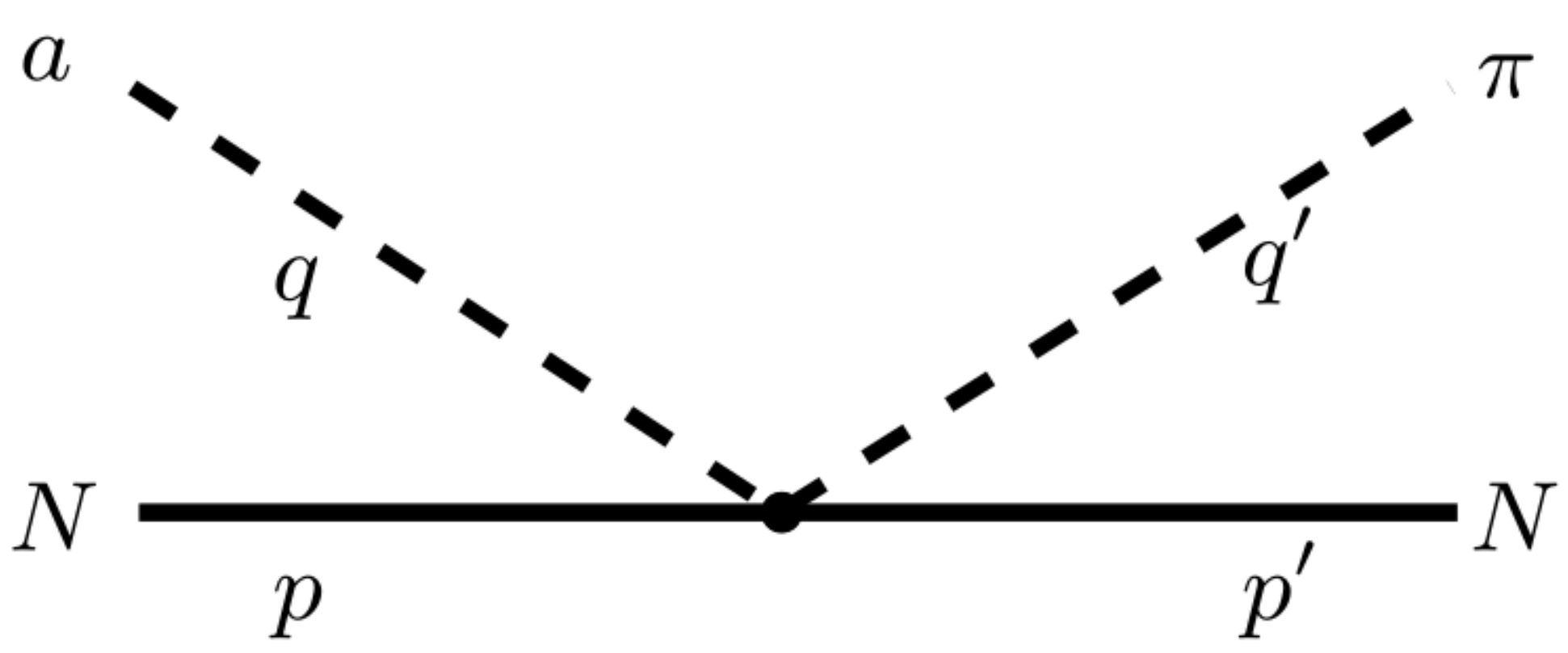}
\label{fig:contact}}\qquad
\subfloat[]{
\includegraphics[width=0.28\textwidth]{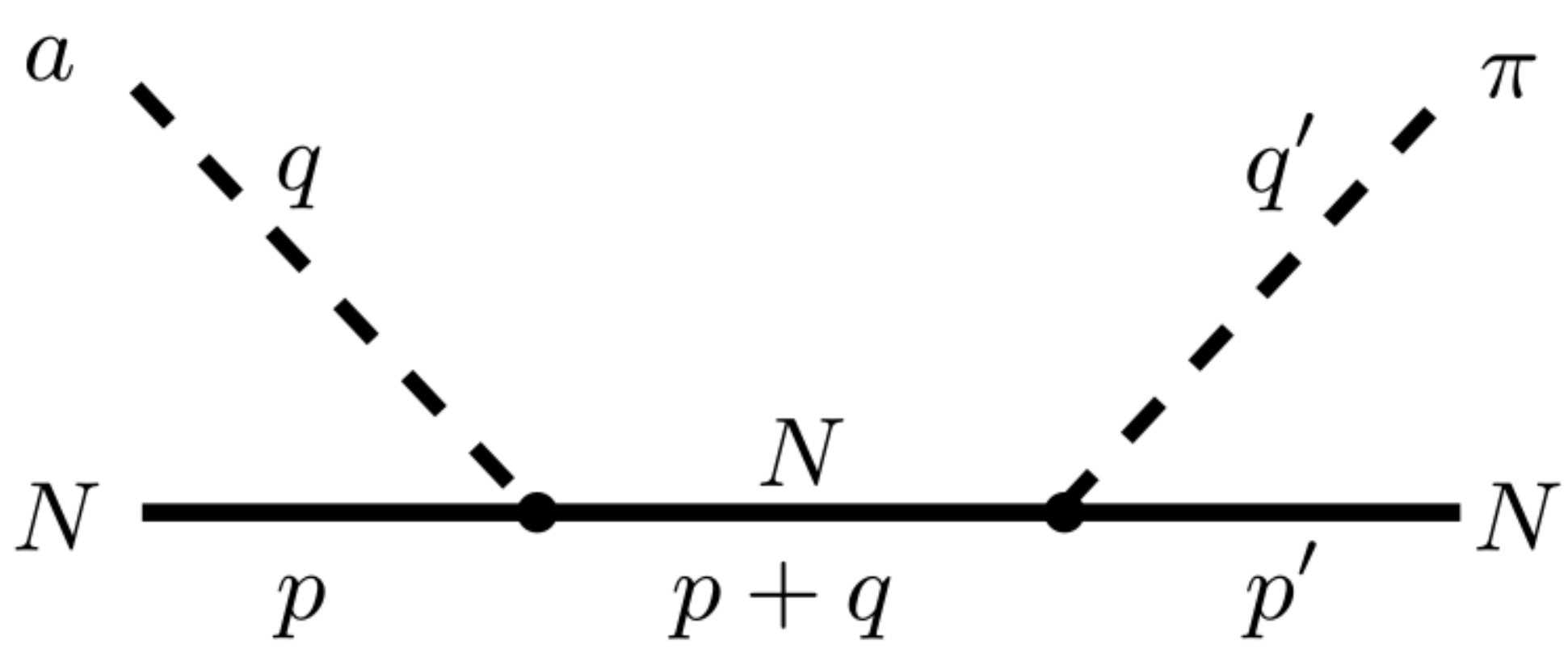}
\label{fig:ns}}\qquad
\subfloat[]{
\includegraphics[width=0.28\textwidth]{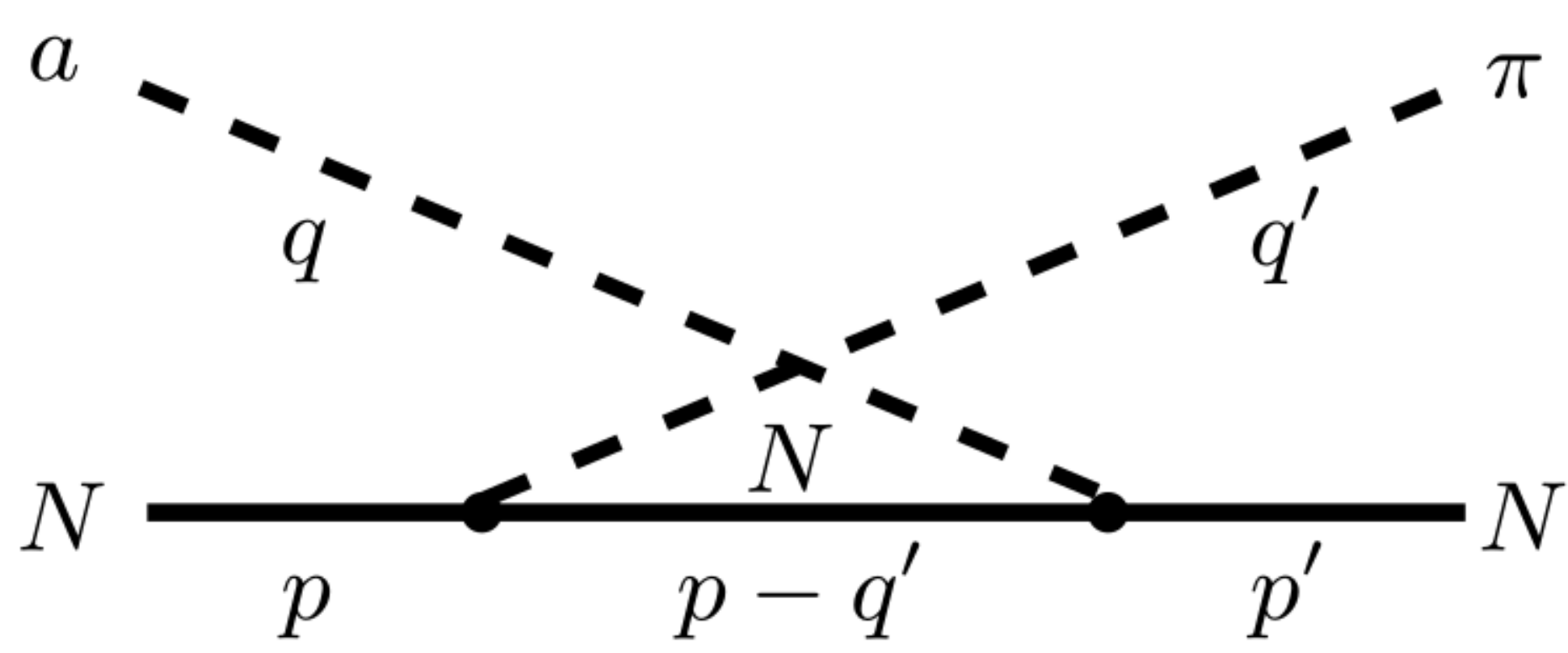}
\label{fig:ncrossed}}
\caption{Tree level contributions to $aN\to \pi N$ without the $\Delta$ intermediate state.}
\label{fig:tree}
\end{figure*}

Having set up the kinematic environment and the theoretical framework, we can now explore 
several contributions to the $aN\to \pi N$ scattering amplitude. We start with the tree-level
contact and Born graphs shown in Fig.~\ref{fig:tree}. The results can be obtained in a rather
straightforward fashion by using the vertices of Eq.~\eqref{eq:vertices} and the $\pi N$ vertex of
the Lagrangian Eq.~\eqref{eq:piNLagrangian}.

The contact interaction, Fig.~\ref{fig:contact}, only gives a contribution to $B^-$ and is free
of any kinematic variable:
\begin{equation}
B^-_\text{\ref{fig:contact}} = \frac{c_{u-d}}{2f_a F_\pi}\,.
\end{equation}
This means that the contact interaction is solely present in the $a p\to \pi^+ n$ and $a n\to \pi^- p$
processes, but absent in any process involving the neutral pion. For the diagrams of
Figures~\ref{fig:ns} and \ref{fig:ncrossed}, one gets
\begin{widetext}
\begin{align}
\begin{split}
A^+_\text{\ref{fig:ns},\ref{fig:ncrossed}} & = \frac{g_A^2 c_{u-d} m_N}{f_a F_\pi} , \\  B^+_\text{\ref{fig:ns},\ref{fig:ncrossed}} (s,t) & = -\frac{g_A^2 c_{u-d} m_N^2}{f_a F_\pi} \left(\frac{1}{s-m_N^2}-\frac{1}{u-m_N^2}\right), \\
A^{3+}_\text{\ref{fig:ns},\ref{fig:ncrossed}}  & = \frac{g_A g_0^i c_i m_N}{f_a F_\pi} , \\  B^{3+}_\text{\ref{fig:ns},\ref{fig:ncrossed}} (s,t) & = - \frac{g_A g_0^i c_i m_N^2}{f_a F_\pi}\left(\frac{1}{s-m_N^2}-\frac{1}{u-m_N^2}\right)  , \\
A^-_\text{\ref{fig:ns},\ref{fig:ncrossed}} & = 0 , \\ B^-_\text{\ref{fig:ns},\ref{fig:ncrossed}} (s,t) & = - \frac{g_A^2 c_{u-d} m_N}{2 f_a F_\pi} \left[1+2m_N\left(\frac{1}{s-m_N^2}+\frac{1}{u-m_N^2}\right) \right],\end{split}
\end{align}
\end{widetext}
where $u$ needs to be understood as $u(s,t)$ via Eq.~\eqref{eq:MandelstamIdentity}. In the appendix, we give a different expression of these
contributions in terms of the axion-nucleon coupling constants $g_{an}$ and $g_{ap}$ for each of the
four possible $aN\to \pi N$ channels. However, for the study of the $P_{33}$ partial wave, it is not
expedient to rewrite them in terms of $g_{an}$ and $g_{ap}$, because after forming the difference
Eq.~\eqref{eq:Tdifference}, one can nicely see that the isoscalar terms $\propto c_i$ stemming from
$a_{\mu,i}^{(s)}$ drop out, leaving only the isospin violating portion $\propto c_{u-d}$ that originates
from $a_\mu$. This fact makes it easy to show (as will be done below) that in the case of the $P_{33}$
partial wave the KSVZ axion can be treated as a special case of the DSFZ axion.

\subsection{Intermediate Delta resonance}\label{ssec:delta}

\begin{figure*}[t]
\subfloat[]{
\includegraphics[width=0.3\textwidth]{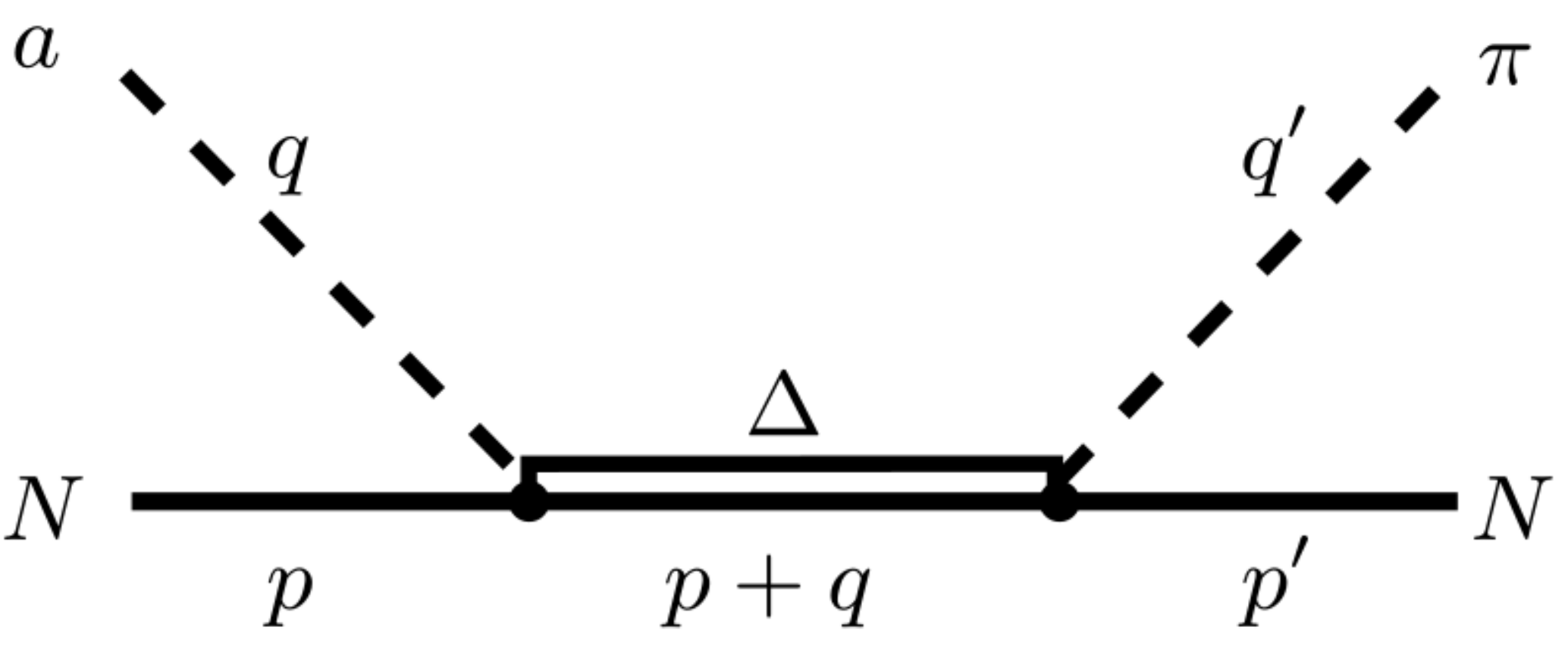}
\label{fig:ds}}\qquad
\subfloat[]{
\includegraphics[width=0.3\textwidth]{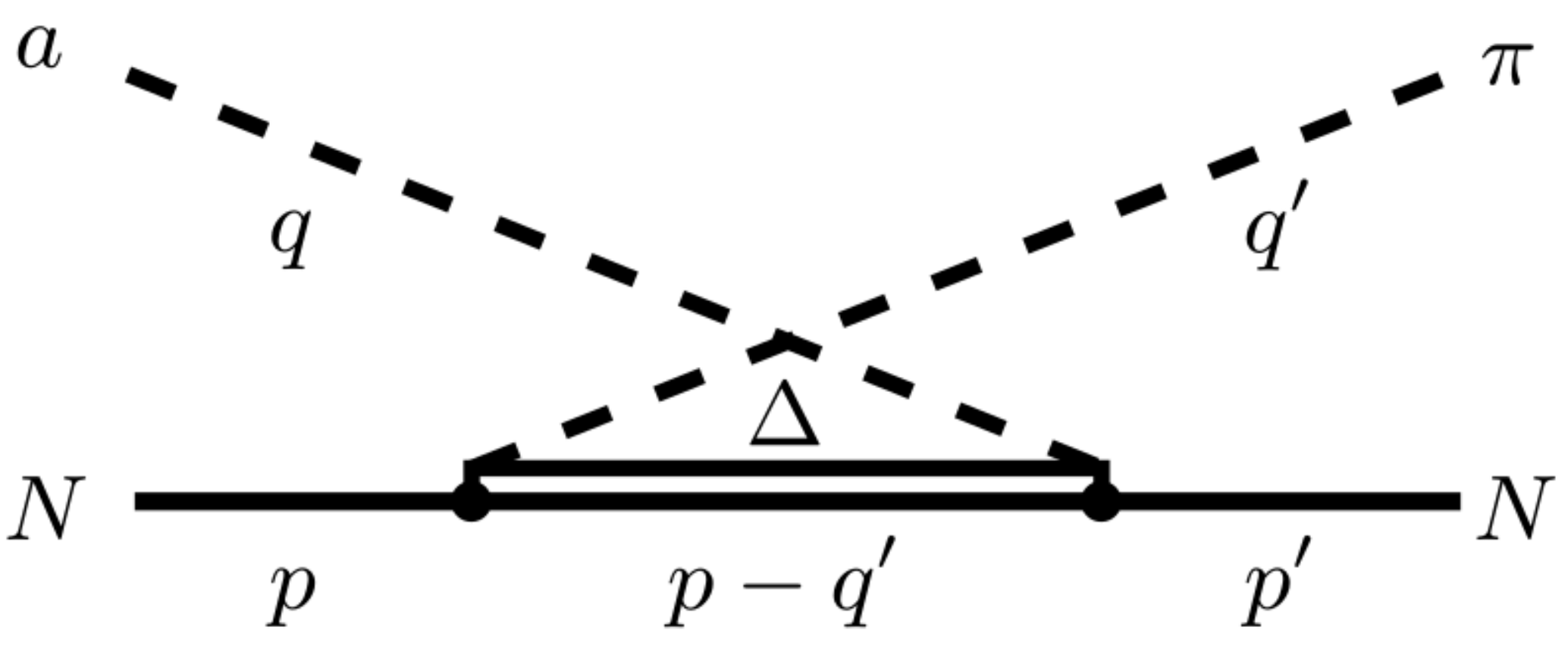}
\label{fig:dcrossed}}
\caption{Tree-level contributions to $aN\to \pi N$ with an intermediate $\Delta$ state.}
\label{fig:delta}
\end{figure*}
Including the $\Delta$ leads to the diagrams shown in Fig.~\ref{fig:delta}. As the two diagrams are
related by crossing, it is convenient to define
\begin{widetext}
\begin{align}\label{eq:Adelta}
\begin{split}
A_\Delta (s,t) = \frac{2 g^2 c_{u-d}}{3f_a F_\pi}\Biggl\{ & \frac{2z_0}{3m_\Delta^2} \left[m_\Delta+\left(m_N+2m_\Delta \right)z_0 \right] \left(s-m_N^2\right)  + \frac{1}{s-\mu_\Delta^2} \biggl[\left(m_N+m_\Delta\right)\left(\frac{1}{2}\left[m_a^2+M_\pi^2-t\right]-\frac{1}{3}\left[s-m_N^2\right]\right) \\
&  -\frac{1}{6m_\Delta^2}\biggl(\left(m_N+m_\Delta\right)\left(\left[m_a^2+M_\pi^2\right]\left[s-m_N^2\right]+m_a^2M_\pi^2\right)  +m_a^2M_\pi^2 m_\Delta+m_N\left(s-m_N^2\right)^2\biggr) \biggr]\Biggr\},
\end{split}
\end{align}
and
\begin{align}\label{eq:Bdelta}
\begin{split}
B_\Delta (s,t) = \frac{2 g^2 c_{u-d}}{3f_a F_\pi}\Biggl\{ & -\frac{z_0}{3m_\Delta^2}\biggl[m_a^2+M_\pi^2+2
\left(s-m_N^2\right)\left(1+z_0\right)  +4m_N m_\Delta\left(1+z_0\right)+4m_N\left(m_N+m_\Delta\right)z_0
\biggr]\\
& + \frac{1}{s-\mu_\Delta^2} \biggl[\frac{1}{2} \left[m_a^2+M_\pi^2-t\right]-\frac{1}{6}m_a^2 +\frac{1}{6m_\Delta}\left(m_N+m_\Delta\right)\left(4m_N m_\Delta - M_\pi^2\right) \\
&  -\frac{1}{6m_\Delta^2}\Bigl[\left(m_a^2+M_\pi^2+2m_N m_\Delta\right)\left(s-m_N^2\right) + m_a^2\left(m_N m_\Delta + M_\pi^2\right) + \left(s-m_N^2\right)^2 \Bigr]\biggr] \Biggr\} .
\end{split}
\end{align}
\end{widetext}
Here the axion mass terms are kept explicitly though they, being tiny for the standard QCD axion models, can be safely neglected.

Then one can combine both diagrams and obtains the following expressions 
\begin{align}
\begin{split}
A_\Delta^+ (s,t) & = A_\Delta (s,t) + A_\Delta (u,t), \\ B_\Delta^+ (s,t)& = B_\Delta (s,t) - B_\Delta (u,t),  \\
A_\Delta^- (s,t)& = -\frac{1}{2} \left[ A_\Delta (s,t) - A_\Delta (u,t) \right], \\ B_\Delta^- (s,t)&
= -\frac{1}{2} \left[B_\Delta (s,t) + B_\Delta (u,t)\right]\,,
\end{split}
\end{align}
where again $u = u(s,t)$. Note that there is no contribution to $T^{3+}$. Equations~\eqref{eq:Adelta}
and~\eqref{eq:Bdelta} have a pole appearing at c.m. energies around the $\Delta$ mass squared.
In order to circumvent any unnecessary subtleties related to 
this, we use a Breit-Wigner propagator with a complex mass squared
\begin{equation}
\mu_\Delta^2 = m_\Delta^2 - i m_\Delta \Gamma_\Delta
\end{equation} 
with $m_\Delta\approx1232$~MeV and $\Gamma_\Delta\approx 117\,\text{MeV}$  the Breit-Wigner mass and width of the $\Delta$ resonance.
A more refined treatment could e.g. be given by including the $\Delta$ self-energy in the complex mass scheme, but that is not required here.

\subsection{Pion rescattering}

\begin{figure}
\centering
\includegraphics[width=0.3\textwidth]{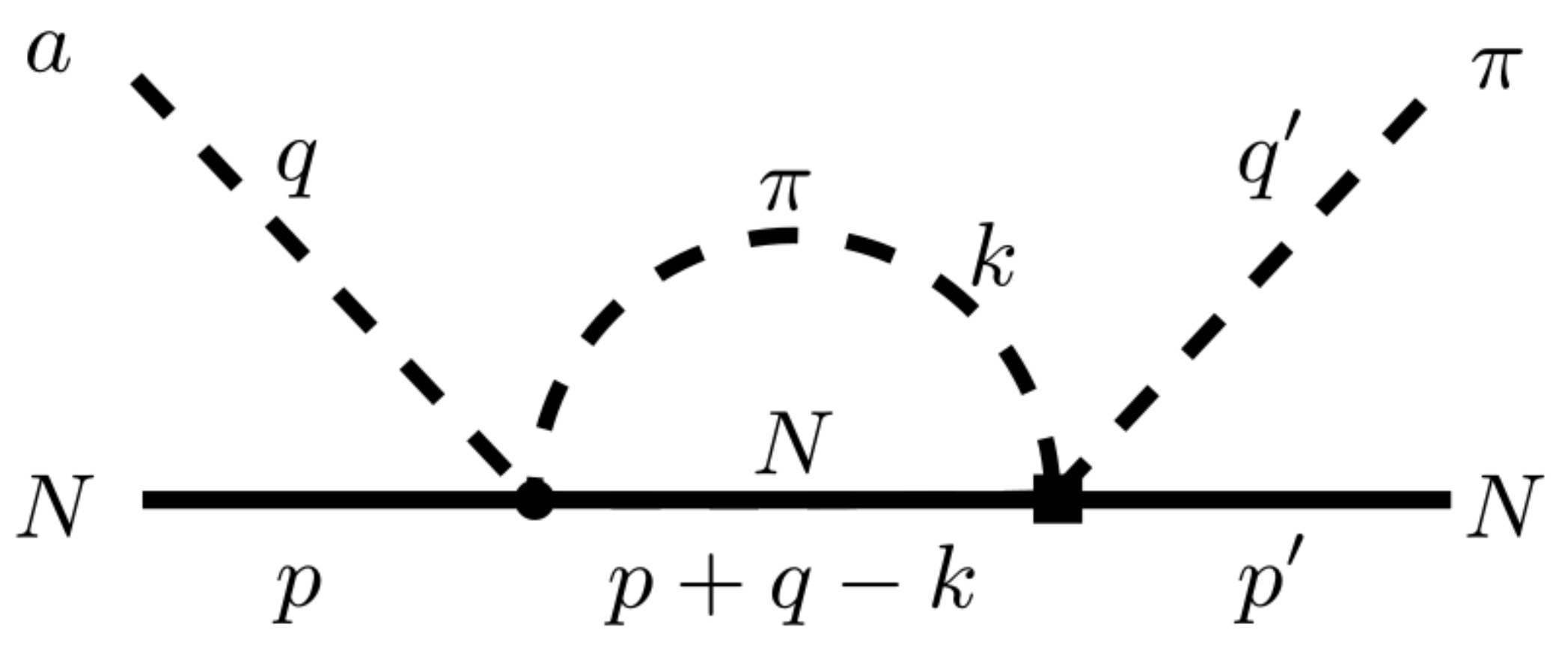}
\caption{The pion rescattering diagram for $aN\to \pi N$.}
\label{fig:rescatter}
\end{figure}

Another sort of diagram that contributes is shown in Fig.~\ref{fig:rescatter}. As in the previous diagrams,
the left (smaller) vertex leads to an axion-pion conversion (so this vertex basically comprises the
contributions of Fig.~\ref{fig:tree}). This pion consequently gets rescattered in the ordinary $\pi N$ scattering.
The latter (larger) vertex treated in a proper way also includes contributions from the $\Delta$ baryon.
As this is an often studied process, we can base the treatment of this diagram on previous results.
In particular, we will adopt the method and results of Refs.~\cite{Meissner:1999vr,Oller:2000fj}.

The diagrams that contribute to the $\pi N$ scattering at this order are basically the same as the ones
discussed in the previous subsections, but with the axion replaced by another pion. Using Eq.~\eqref{eq:piNiso}
and Eq.~\eqref{eq:lprojection} leads to a projection to the $P_{33}$ partial wave. We do not repeat the results
for the diagrams here, they can be found in Ref.~\cite{Meissner:1999vr}, Eqs.~(3.4) and~(3.9).\footnote{As
the diagrams of Fig.~\ref{fig:delta} and the corresponding diagrams of $\pi N$ scattering are basically the
same, the result of Ref.~\cite{Meissner:1999vr} can also be found by replacing $m_a\to M_\pi$ in Eqs.~\eqref{eq:Adelta}
and~\eqref{eq:Bdelta}, adjusting of course  the coupling constants in front.} Moreover, we also adopt the results
for the renormalized chiral pion loops obtained in heavy baryon CHPT (HBCHPT) from \cite{Fettes:1998ud}
(which are needed for the unitarization, see below).

As the $\pi N$ scattering above threshold and below the appearance of inelastic reactions fulfills the unitarity
relation (here $W=\sqrt{s}$, the c.m. energy)
\begin{equation}\label{eq:unitarity}
\operatorname{Im}T_{l\pm}^I (W) = \frac{|\mathbf{q}|}{8\pi W} |T_{l\pm}^I (W)|^2,
\end{equation}
the pole in the $\Delta$ propagator can be treated in a more systematic way than the one we used in
Sec.~\ref{ssec:delta} for the $aN\to\pi N$ reaction. In particular, a suitable unitarization technique
can be used to restore unitarity which is otherwise only fulfilled perturbatively in CHPT. The method
used in Ref.~\cite{Meissner:1999vr} is the $N/D$ method~\cite{Chew:1960iv,Oller:2020guq}. In a nutshell,
it is based on the observation that the unitarity relation leads to a right-hand cut in the partial
wave $T$-matrix such that one can write down a dispersion relation for the inverse amplitude with some
extra terms which are free of any right-hand cuts. These can be matched to the amplitudes
obtained from CHPT. This effectively corresponds to a resummation of the relevant diagrams. Possible
double counting can be avoided by the matching procedure discussed in Ref.~\cite{Oller:2000fj}.
The integral of the dispersion relation can be performed analytically and is basically given by the
known two-point loop function involving one pion and one nucleon,
\begin{align}\label{eq:loopfct}
g(s) =&\, \frac{1}{16\pi^2} \left\{a_0(\mu) + \left(1 - \frac{w}{m_N}\right) \log\left(\frac{M_\pi^2}{\mu^2}\right) \right. \nonumber\\
& \left. - x_+ \log\left(\frac{x_+ - 1}{x_+}\right) - x_- \log\left(\frac{x_- - 1}{x_-}\right)\right\},
\end{align}
at a renormalization scale $\mu$. We take $\mu$ as the nucleon mass, and any change in $\mu$ can be reabsorbed by the subtraction constant $a_0(\mu)$. Furthermore, $w$ is the c.m. pion energy and
\begin{equation}
x_\pm = \frac{s + m_N^2 - M_\pi^2}{2s} \pm \frac{1}{2 s} \sqrt{\lambda (s, m_N^2, M_\pi^2)}\,.
\end{equation}
The matching procedure for unitarizing the leading one-loop $\mathcal{O}(p^3)$ amplitude with the $\Delta$ resonance leads to
\begin{widetext} 
\begin{equation}
T^{I,l\pm}_{\pi N} = \frac{1}{\left(T^{I,l\pm}_\text{tree}+T^{I,l\pm}_\text{loop}+T^{I,l\pm}_\Delta
+\frac{2\sqrt{s}}{E_\mathbf{p}+m_N} \left(T^{I,l\pm}_\text{LO}\right)^2 g(s)\right)^{-1}+g(s)},
\end{equation}
\end{widetext}
which indeed fulfills Eq.~\eqref{eq:unitarity}. Note that the tree contributions $T^{I,l\pm}_\text{tree}$
and $T^{I,l\pm}_\Delta$ for the resonance are taken as being the full relativistic ones, whereas
$T^{I,l\pm}_\text{LO}$ is only the very leading order HBCHPT amplitude.

\begin{figure}
\centering
\includegraphics[width=\linewidth]{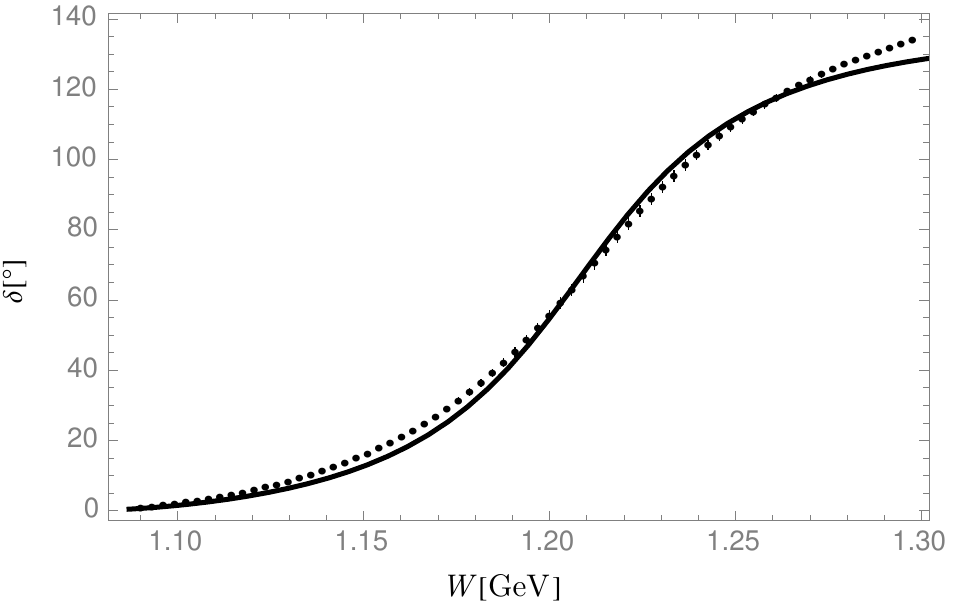}
\caption{The $\pi N$ phase shift $\delta$ in the $P_{33}$ channel (solid line) fitted to the results
of the Roy--Steiner analysis (dots) from~\cite{Hoferichter:2015hva}.}
\label{fig:fitP33}
\end{figure}

Returning to  pion axioproduction, we performed a full reanalysis of the phase shift $\delta_{l\pm}^I$ defined by
\begin{equation}
T_{l\pm}^I (W) = \frac{8\pi W}{|\mathbf{q}|}\exp(i\delta_{l\pm}^I)\sin(\delta_{l\pm}^I)
\end{equation}
in order to use these results for the rescattering diagram and in order to determine accurate values
for the coupling constants $g$ and $z_0$ of Eq.~\eqref{eq:deltainteractionlagr} and $a_0$ of Eq.~\eqref{eq:loopfct}.
The latter goal is achieved by fitting the resulting $P_{33}$ phase shift to the results of the Roy--Steiner
analysis of the $\pi N$ scattering~\cite{Hoferichter:2015hva}. As input values we used the isospin-averaged nucleon mass
$m_N = (m_n+m_p)/2 = 938.92\,\text{MeV}$, the isospin-averaged pion mass $M_\pi = 138.03\,\text{MeV}$,
$F_\pi=92.4\,\text{MeV}$, and $m_\Delta = 1232\,\text{MeV}$. The value of $g_A$ is discussed below when it
comes to the determination of the axion-baryon couplings in Sec.~\ref{sec:result}. The fit to the phase shift values
of $W\lesssim 1.3\,\text{GeV}$ yields
\begin{equation}
g = 1.249(16),\quad z_0 = -0.21(56),\quad a_0 = -0.959(12)
\end{equation}
and is shown in Fig.~\ref{fig:fitP33}.

Finally, the rescattering diagram is evaluated by
\begin{align}
\begin{split}\label{eq:rescatter}
T^{33}_{\rm rescatt.} (s) = & \Bigl( T^{33,\text{tree}}_{ap\to\pi^0 p} (s) g(s,M_{\pi^0}) \\ & \quad - \frac{1}{\sqrt{2}}
T^{33,\text{tree}}_{ap\to\pi^+ n}(s) g(s,M_{\pi^+}) \Bigr) T^{33}_{\pi N}(s)\,,
\end{split}
\end{align}
where $g(s,M_{\pi^{+,0}})$ is the pion-nucleon loop function Eq.~\eqref{eq:loopfct} with the meson (nucleon) mass being the
charged and neutral pion (neutron and proton) mass, respectively. $T^{33,\text{tree}}_{aN\to\pi N}$ denotes the partial wave
projected amplitudes of Sec.~\ref{ssec:treeaN} and $T^{33}_{\pi N}$ the unitarized $P_{33}$ partial wave
amplitude just discussed. We consider the usage of the latter appropriate even though it is derived
using on-shell kinematics, as the off-shell effects are certainly subleading~\cite{Mai:2012dt}.

The expression in the parentheses is the proper way of getting the isospin violating part of the amplitude,
as discussed in Sec.~\ref{ssec:isospin}, see Eq.~\eqref{eq:Tdifference2}, i.e. by taking the difference
of the amplitudes and the difference in the charged and neutral pion/nucleon masses. If one neglects this mass
difference, one might as well use Eq.~\eqref{eq:Tdifference} instead yielding
\begin{equation}\label{eq:rescatter2}
T^{33}_{\rm rescatt.}(s) =T^{33,\text{tree}}_{aN\to\pi N}(s) g(s,M_{\pi}) T^{33}_{\pi N}(s),
\end{equation}
where $T^{33,\text{tree}}_{aN\to\pi N}$ now is the $j=\tfrac{3}{2}$ projection of $T^{I=3/2}$, Eq.~\eqref{eq:Tdifference},
and $M_\pi$ is the isospin averaged pion mass. In fact, this latter approximation gives an average deviation of
only $\lesssim 2\%$ in comparison to Eq.~\eqref{eq:rescatter}, which is valid for the KSVZ axion and the DFSZ axion
at $\sin^2\beta\lesssim 0.95$. Only for values $\sin^2\beta\to 1$ the deviation becomes more pronounced
at c.m. energies $W \gtrsim 1.15\,\text{GeV}$ reaching a maximum of about $15\%$. This means
that Eq.~\eqref{eq:rescatter2} is a very good approximation for the vast majority of cases.

\section{Results}\label{sec:result}

Let us take the last result of the previous section as a starting point for the discussion of the overall
results of this study. If one indeed takes the approximation of equal pion/nucleon masses, then Eq.~\eqref{eq:isoprojection}
causes that any dependence on the isoscalar couplings $g_0^i c_i$ is canceled and any dependence on the
axion-nucleon coupling $g_{aN}$ in this process reduces to a dependence on $g_A c_{u-d}$ only, which reflects that this part of the coupling enforces the isospin violation needed to enable the $\Delta$
resonance appearance. For the same reason, the diagrams of Fig.~\ref{fig:delta} with the explicit $\Delta$
solely depend on $c_{u-d}$ and not on the $c_i$'s. This has two consequences: First, the total amplitude for
the DFSZ axion with $\sin^2\beta$ in the interval $[0,1]$ and that for the KSVZ model will be $\propto c_{u-d}$; they 
have entirely the same shape, and only the magnitude changes as $\sin^2\beta$ is varied. Second, as $c_{u-d}$
only depends on the difference $X_u-X_d$, one can easily determine a value for $\sin^2\beta$ such that
$c_{u-d}^\text{DFSZ}(\sin^2\beta)=c_{u-d}^\text{KSVZ}$, which is accomplished when $\sin^2\beta=\tfrac{1}{2}$
(see Eq.~\eqref{eq:couplignconstantsmodeldepending}). The $aN\to\pi N$ scattering amplitude in the
$P_{33}$ channel for the KSVZ axion is hence exactly the same as that for the DFSZ axion at $\sin^2\beta=\tfrac{1}{2}$.
As the deviation from this approximation is only $\lesssim 2\%$, this remains basically true even if
one considers Eq.~\eqref{eq:rescatter} instead of Eq.~\eqref{eq:rescatter2}.

For the calculation of the final scattering amplitude, we make use of the nucleon matrix elements in order
to determine the isovector and isoscalar axial-vector couplings
\begin{eqnarray}
g_A &=& \Delta u-\Delta d\,, \nonumber \\
g_0^{u+d} &=& \Delta u+\Delta d\,, \\
g_0^q &=& \Delta q\,,\text{ for } q=s,c,b,t\ , \nonumber
\end{eqnarray}
where $s^\mu\Delta q=\langle p | \bar{q}\gamma^\mu\gamma_5 q|p\rangle$, with $s^\mu$ the spin of the proton.
Of course, for the approximation discussed in the previous paragraph, only the value of $g_A$ is of
interest. For these matrix elements and $z$ and $w$ appearing in Eq.~\eqref{eq:couplignconstantsmodeldepending},
we take the recent values from Ref.~\cite{Aoki:2021kgd},
\begin{equation}
\begin{array}{rlrlrl}
\Delta u & = 0.847(50) , \\ \Delta d & = -0.407(34) , \\ \Delta s & = -0.035(13)  , \\
z & = 0.485(19) , \\ w & = 0.025(1)  , \\
\end{array}
\end{equation}
and ignore $\Delta q$ for $q=c,b,t$. 

\begin{figure*}[t]
\centering
\includegraphics[width=\textwidth]{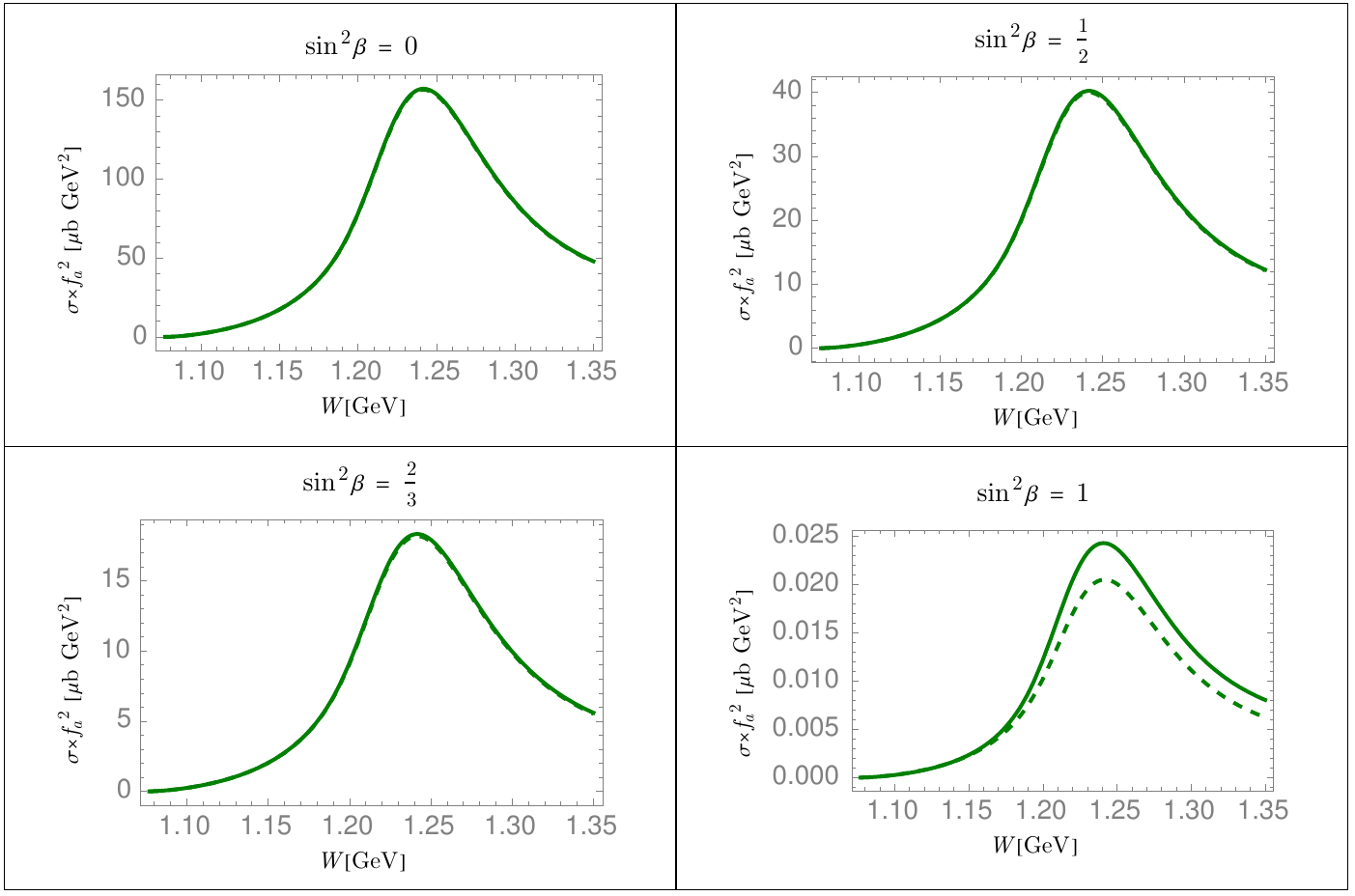}
\caption{The $a N\to\pi N$ partial wave cross section of the $P_{33}$ channel versus the c.m. energy $W$ for
the DFSZ axion at different values of $\sin^2\beta$ and the KSVZ axion (which corresponds to $\sin^2\beta
= \frac{1}{2}$, see main text). The dashed curve corresponds to the approximated case based on
Eq.~\eqref{eq:rescatter2}.
}
\label{fig:aNscatt}
\end{figure*}

In Fig.~\ref{fig:aNscatt}, we show the partial wave cross sections $\sigma^{33}_{aN\to\pi N}$ consisting of all
the contributions discussed in Sec.~\ref{sec:amplitudes}, for both with the approximation
Eq.~\eqref{eq:rescatter2} and without it. Actually, the cross sections are multiplied by the
factor $f_a^2$ in order to get rid of the unknown prefactor $1/f_a^2$. This unknown quantity
also appears implicitly in the terms containing the axion mass in Eq.~\eqref{eq:Adelta} and~\eqref{eq:Bdelta},
but has practically no effect as the axion mass $\propto 1/f_a$ can safely be neglected for the
typical QCD axion window, Eq.~\eqref{eq:window}. However, this prefactor has to be kept in mind when
considering the strength of the amplitudes in Fig.~\ref{fig:aNscatt}.

As anticipated, the curves for different values of $\sin^2\beta$ are identical up to the order of magnitude.
As the absolute value of $c_{u-d}$ is a linearly decreasing function of $\sin^2\beta$, the magnitude
steadily decreases, which makes a DFSZ axion with $\sin^2\beta\to 1$ the most unfavorable candidate for
detection. As expected from the description at the end of the previous section, there is almost no
visual deviation of the curves with approximation Eq.~\eqref{eq:rescatter2} and without it. Only
for $\sin^2\beta=1$ this deviation becomes recognizable but is still a minor effect. Note that the figures for the limit values of $\sin^2\beta = \{0,1\}$ are given rather for illustrative purposes, as in realistic DFSZ models perturbative constraints from the heavy quark Yukawa couplings yield an allowed range $[0.25,170]$ for $\cot\beta$ \cite{DiLuzio:2020wdo} corresponding to approximately $\sin^2\beta\in [0.00,0.94]$.

As a result, there is indeed a considerable enhancement of the $P_{33}$ partial wave cross section in the region of
the $\Delta$ resonance, but this enhancement is considerably weaker than previously assumed by Carenza et
al.~\cite{Carenza:2020cis}, who estimated the cross section via $f_a^2 \sigma_{aN\to\pi N} \approx
F_\pi^2 \sigma_{\pi N\to \pi N}$ taking a value of $100\,\mathrm{mb}$ for $\sigma_{\pi N\to \pi N}$.\footnote{The $\pi^+p$ elastic cross section is the largest around the $\Delta$ resonance region, of about 100~mb~\cite{ParticleDataGroup:2020ssz}.} This
estimation suggests a peak value $f_a^2 \sigma_{aN\to\pi N} \approx 1\,\mathrm{mb}\,\mathrm{GeV}^2$.
The discrepancy between the results of Fig.~\ref{fig:aNscatt} and such estimation can be
explained by the fact that ${aN\to\pi N}$ in the $\Delta$ sector is primarily an isospin breaking process,
which always comes with an extra suppression. It is worthwhile to notice that the suppression of the isospin breaking here, characterized by the factor $(1-z)/(1+z) = (m_d-m_u)/(m_d+m_u) \approx 0.34$ for the model-independent part of $c_{u-d}$ (see Eq.~\eqref{eq:cuds}), is much milder than that for usual isospin breaking in hadronic processes, characterized by $(m_d-m_u)/m_s$ or $(m_d-m_u)/\Lambda_\text{QCD}$. Thus, the results given in Ref.~\cite{Carenza:2020cis} are
to be multiplied by a factor $10^{-1}$ to $10^{-5}$, depending on the value of the model-dependent factor $\sin^2\beta$, that is a suppression by at least one order of magnitude.
Then the number of pions produced via $aN\to \pi N$  through the $\Delta$ resonance in a megaton water Cherenkov detector will be at most $\mathcal{O}(100)$ using the axion luminosity estimated in Ref.~\cite{Carenza:2020cis} for axions emitted from a supernova at 1 kiloparsec.

\section{Summary}
In this study we presented an analysis of the pion axioproduction $aN\to\pi N$ with an
intermediate $\Delta$ resonance. We included the $\Delta$ resonance in two different ways: First, we
used the chiral interaction Lagrangian for the $\Delta$ to bring the axion explicitly into contact with
the resonance, and, second, we used the well-known results of $\pi N$ elastic scattering with $\Delta$
to include it implicitly in the form of the rescattering diagram in Fig.~\ref{fig:rescatter}. 

As the $\Delta$ is a spin-$\tfrac{3}{2}$ and isospin-$\tfrac{3}{2}$ particle, it shows its full leverage
effect in the $P_{33}$ partial wave, which is why we concentrated on a study of this particular partial wave.
For the same reason, this interaction is essentially an isospin violating process, as the axion is an isosinglet.
We have shown that an approximation that concentrates on this isospin violation and that neglects any
isoscalar coupling, while at the same time ignoring the pion and nucleon isospin mass splittings, is still very accurate
unless $\sin^2\beta$ approaches $1$ in the DFSZ axion model (where it still gives quite good results). In
this way, it is shown that the partial wave amplitude for the  KSVZ axion equals that for the DFSZ axion at
$\sin^2\beta=\tfrac{1}{2}$.

Finally, the enhancement of the amplitude anticipated by Carenza et al.~\cite{Carenza:2020cis} is indeed
present in the region of the $\Delta$ resonance, although it is considerably weaker than their naive estimation by at least an order of magnitude. This is basically a consequence of the isospin breaking suppression which though is much milder than that for usual isospin breaking hadronic processes. 
Therefore, it might be interesting to check whether other isospin-1/2 resonances such as the $N^*(1440)$ Roper resonance would provide an additional enhancement of the $aN\to\pi N$ cross section, as
it is accessible without isospin breaking. The next step would hence be to investigate the impact of
such resonances on the $aN\to\pi N$ reaction and consequently on the axion production in stellar objects, and
whether this might be exploited to give fresh perspectives on experimental axion searches. 

\appendix*
\section{Amplitudes of the leading order tree graphs}\label{app:1}

In this appendix, we give the full expression of the leading order tree graph amplitudes of the four $aN\to\pi N$ channels. Using the abbreviation
\begin{equation}
R_\mu = \frac{1}{2} \left(q_\mu + q_\mu^\prime\right) ,
\end{equation}
the contact contribution reads
\begin{align}
T^\text{\ref{fig:contact}}_{aN\to\pi^0 N} & = \phantom{-} 0  , \\
T^\text{\ref{fig:contact}}_{ap\to\pi^+ n} & = \phantom{-}\frac{c_{u-d}}{\sqrt{2}f_aF_\pi}\bar{u}(p^\prime)\,  \slashed{R}\, u(p) , \\
T^\text{\ref{fig:contact}}_{an\to\pi^- p} & = -\frac{c_{u-d}}{\sqrt{2}f_aF_\pi}\bar{u}(p^\prime) \, \slashed{R}\, u(p) ,
\end{align}
where the superscript refers to Fig.~\ref{fig:tree}. For the other two diagrams Fig.~\ref{fig:ns} and ~\ref{fig:ncrossed} one gets
\begin{widetext}
\begin{align}
T^\text{\ref{fig:ns},\ref{fig:ncrossed}}_{ap\to\pi^0 p}& =  \phantom{-} \frac{g_A m_N g_{ap}}{f_a F_\pi} \bar{u}(p^\prime) \Biggl\{ 1-m_N \left(\frac{1}{s-m_N^2}-\frac{1}{u-m_N^2} \right) \slashed{R}\Biggr\} u(p), \\ 
T^\text{\ref{fig:ns},\ref{fig:ncrossed}}_{an\to\pi^0 n}& =  - \frac{g_A m_N g_{an}}{f_a F_\pi} \bar{u}(p^\prime) \Biggl\{ 1-m_N \left(\frac{1}{s-m_N^2}-\frac{1}{u-m_N^2} \right)\slashed{R}\Biggr\} u(p), \\
T^\text{\ref{fig:ns},\ref{fig:ncrossed}}_{ap\to\pi^+ n} &= \frac{g_A}{\sqrt{2} f_a F_\pi} \bar{u}(p^\prime) \Biggl\{ 2m_N g_0^ic_i -\left[  g_A c_{u-d} + 2m_N^2 \left(\frac{g_{ap}}{s-m_N^2} -\frac{g_{an}}{u-m_N^2}\right)\right]\slashed{R}\Biggr\} u(p),\\
T^\text{\ref{fig:ns},\ref{fig:ncrossed}}_{an\to\pi^- p} & =  \frac{g_A}{\sqrt{2} f_a F_\pi} \bar{u}(p^\prime) \Biggl\{ 2m_N g_0^ic_i  + \left[ g_A c_{u-d} - 2m_N^2 \left(\frac{g_{an}}{s-m_N^2} -\frac{g_{ap}}{u-m_N^2}\right)\right]\slashed{R}\Biggr\} u(p) .
\end{align}
\end{widetext}
Here, $g_{an}$ and $g_{ap}$ are the usual axion-nucleon couplings given in Eq.~\eqref{eq:aNcoupling}.

\begin{acknowledgments}
We thank Alessandro Mirizzi, Pierluca Carenza, and Maurizio Giannotti for directing our attention to this topic.
Jacobo Ruiz de Elvira kindly provided us with the Roy--Steiner equation analysis data.
This work is supported in part by the Deutsche Forschungsgemeinschaft (DFG)
and the National Natural Science Foundation of China (NSFC) through the funds provided to the
Sino-German Collaborative Research Center ``Symmetries and the Emergence of Structure in QCD"
(NSFC Grant No. 12070131001, DFG
Project-ID 196253076 -- TRR 110), by the NSFC under Grants No.~12125507, No.~11835015 and No.~12047503, by the Key Research Program of the Chinese
Academy of Sciences (CAS) under Grant No.~XDPB15,  by the CAS 
President's International Fellowship
Initiative (PIFI) (Grant No.~2018DM0034), by the VolkswagenStiftung (Grant No. 93562), and by the
EU (STRONG2020). 
\end{acknowledgments}

%

\end{document}